\documentclass{article}

\usepackage{arxiv}

\usepackage[utf8]{inputenc} 
\usepackage[T1]{fontenc}    
\usepackage{hyperref}       
\usepackage{url}            
\usepackage{booktabs}       
\usepackage{amsfonts}       
\usepackage{nicefrac}       
\usepackage{microtype}      
\usepackage{lipsum}		
\usepackage{graphicx}
\usepackage{doi}
\usepackage[T1]{fontenc}
\usepackage{booktabs}
\usepackage{subscript}
\usepackage{fixltx2e}

\usepackage{graphicx,booktabs} 
\usepackage{url,hyperref,lineno,microtype,subcaption}
\usepackage{float}
\usepackage{verbatim} 
\usepackage{multirow}
\usepackage{booktabs}
\usepackage{textcomp}
\usepackage{siunitx}
\usepackage{amsmath}
 \usepackage{enumitem}
 \usepackage{makecell}
 \usepackage{siunitx}
 \usepackage{amsfonts}
\usepackage{gensymb}
\usepackage[table,xcdraw]{xcolor}
\usepackage{url}

\title{ Sonified distance in sensory substitution does not always improve localization: comparison with a 2D and 3D handheld device.}

\author{\hspace{1mm}Louis Commère, Jean Roaut \\
    Department of Electrical Engineering\\
	Université de Sherbrooke\\
	2500 Boulevard de l’Université, Sherbrooke, QC J1K 2R1, Canada \\
	\texttt{\{louis.commere,jean.rouat\}@usherbrooke.ca} \\
}

\renewcommand{\shorttitle}{\textit{arXiv} Template}

\hypersetup{
pdftitle={A template for the arxiv style},
pdfsubject={q-bio.NC, q-bio.QM},
pdfauthor={David S.~Hippocampus, Elias D.~Striatum},
pdfkeywords={First keyword, Second keyword, More},
}

\begin{document}
\maketitle

\begin{abstract}
Early visual to auditory substitution devices encode 2D monocular images into sounds while more recent devices use distance information from 3D sensors. 
 This  study assesses whether the addition of sound-encoded distance in recent systems helps to convey the ``where'' information. 
 This is important to the design of new sensory substitution devices. 
 We conducted experiments for 
  object localization and navigation tasks with a handheld visual to audio substitution system. It comprises 2D and 3D modes. Both encode in real-time the position of objects in images captured by a camera. 
 The 3D mode encodes in addition the distance between the system and the object.
Experiments have been conducted with 16 blindfolded sighted participants. 
 For the localization, participants were quicker to understand the scene with the 3D mode that encodes distances. On the other hand, with the 2D only mode, they were able to compensate for the lack of distance encoding after a small training. For the navigation, participants were as good with the 2D only mode than with the 3D mode encoding distance.  
\end{abstract}


\section{Introduction}
Sensory 
substitution is a mechanism 
by which one sense can stimulate and consolidate brain areas usually associated with 
 another sense, given the extremely plastic nature of the brain~\cite{bach1987brain,kokovayNeuron2008}. 
Initial research conducted by P. Bach-Y Rita~\cite{bachYRita1969} established that the 
back can be used to mediate visual stimulus to the brain. Then, he showed that trained blind people can 
navigate and ``visually perceive'' 
features of the environment via electric pulses on the tongue that encode camera images~\cite{bachYRitaTrends2003,ptitoBrain2005}.
The substitution from vision to audition has also been studied and Sensory Substitution Devices (SSDs) have been designed for that purpose~\cite{csapo2015survey}. We are interested in the potential of visual to auditory SSD to convey 3D spatial information through sounds. 

Access to 3D information is necessary to perform everyday life tasks such as detecting obstacles while navigating or localize and grasp objects. Three dimensional information is however mainly provided with vision and is acquired through monocular or binocular vision. Monocular cues include, for example, the relative size (far objects are perceived smaller) or the motion parallax (for a moving observer, nearby objects appear to be moving faster than distant objects). Binocular cues  include, for example, binocular disparity (position difference of an object on the retina seen by the left and the right eyes). A complete review of visual cues for 3D perception is available in the work of Palmer~\textit{et al.}~\cite{palmer1999vision}. 

In this work, we study the effectiveness of distance encoding into sounds with a handheld visual to auditory SSDs for localization and navigation tasks. As far as we know, no previous research has quantified the potential advantages of using audio encoded distance in such a context. We review below the main audio to vision substitution devices 
 which have been successfully tested for localization and navigation. We classify these systems into two categories based on whether or not they use distance information. Then, we propose a protocol for comparing them. 

Early visual to auditory SSDs convert 2D monocular images from a camera into sounds
 ~\cite{Meijer1992,Capelle1998,Auvray2006,abboud2014eyemusic,ambard2015mobile}. We refer to these as ``2D systems'' (i.e. without distance encoded into sounds). 
Studies have shown that it was possible to infer the depth with such systems despite not having 3D sensors.
Ward~\textit{et al.}~\cite{ward2010} conducted a phenomenology study with two long term blind users of \emph{the vOICe}~\cite{Meijer1992}. \emph{The vOICe} is the first 
known vision to audition device 
which turns 2D monocular greyscale images into sound 
with a mapping of the positions and greyscale values of pixels to frequency and amplitude short-time sine tones. 
After months of immersive use of the system, the two blind people reported being able to evaluate distances. 
Auvray~\textit~{et al.}~\cite{auvray2007learning} also conducted experiments with~\emph{the vOICe} and showed that blindfolded participants were able to localize objects within a 3D environment. 
%
Renier and Volder~\cite{Renier2010} conducted localization experiments with 
early blind and sighted blindfolded participants using the Prosthesis for Substitution of Vision by Audition (\emph{PSVA})~\cite{Capelle1998}. The~\emph{PSVA} holistically encodes the position of pixels from 2D monocular images into frequencies of sine tones with a greater resolution in the center of the image to mimic the fovea. Participants had to estimate the position of cubes placed on a table. The cubes were then removed from the table and participants were able to correctly 
put them back at their initial position. 
People were able to use audio encoded monocular cues of visual to auditory SSDs to perceive distance. 

%


A new generation of vision to audition SSDs takes advantage of the accessibility of 3D sensors to directly encode distance information with sound. We refer to these as ``3D systems'' (i.e. with distance encoded into sounds). 
Some sonify the 3D raw data and the user has to interpret the sounds to understand the environment 
~\cite{Brock2013,stoll2015navigating,Skulimowski2018}. 
Other provide higher level information to the user. 
Several systems algorithmically find the free path to allow users to move safely without hitting obstacles~\cite{yang2016expanding,yang2018long,aladren2016navigation}.
Kayukawa~\textit{et al.}~\cite{Kayukawa2019} proposed a device to detect pedestrians who would collide with the blind user. 
Recent systems also combine object recognition with machine learning and navigation aid~\cite{Bai2019,Li2020}. 

2D systems are 
cheaper 
and distance can be inferred with monocular cues. With 3D ones, the distance is given directly to the user and does not need to be inferred. However, this additional information might overload the hearing system which can confuse users~\cite{kristjansson2016designing}. 

 To our knowledge, no studies have been conducted on the comparison between 2D and 3D handheld devices. Therefore, we designed a protocol to compare the behaviors and performance of individuals using a 2D and a 3D version of a vision to audition SSD.

The protocol comprises localization and navigation experiments with a 2D and 3D mode of the same handheld vision to audition SSD. One mode encodes distance and the other does not. Both rely on the same approach by detecting and encoding into sounds the position of objects in 2D camera images. With the 3D mode, distance between the object and the system is also encoded with an additional sound parameter. Both modes are real-time. 
Our experimental protocol is made of 2 tasks. During the first task, participants became familiar with the device and had to locate, by pointing with the index finger, 3 objects on a table. For the second task, they had to navigate through a corridor while avoiding obstacles randomly placed on the floor. The localization performance is measured with the pointing error distance and the time taken
to find the 3 objects. Navigation performance is evaluated with the number of objects that participants did not detect on their path and the time they took to complete courses. In the end of the experiment, participants had to fill up a questionnaire. 

Experiments were conducted with 16 sighted blindfolded participants. 
 All were able to effectively localize objects and navigate after a short training time (approximately 10 min) with both 2D and 3D modes. At first, they were better at localization with the 3D mode. Then, with learning, similar performance for objects localization were observed with both modes. For navigation, participants were as good with the 2D mode then with the 3D mode. They improved across the courses. 
 Nevertheless, the questionnaire showed that participants generally felt more comfortable with the 3D mode.

\section{Material and methods}
\label{sec:method}
\subsection{Sensory substitution system}
\label{sec:systemDescription}
\begin{figure}[htb] 
   \centering
   \includegraphics[width=0.35\textwidth]{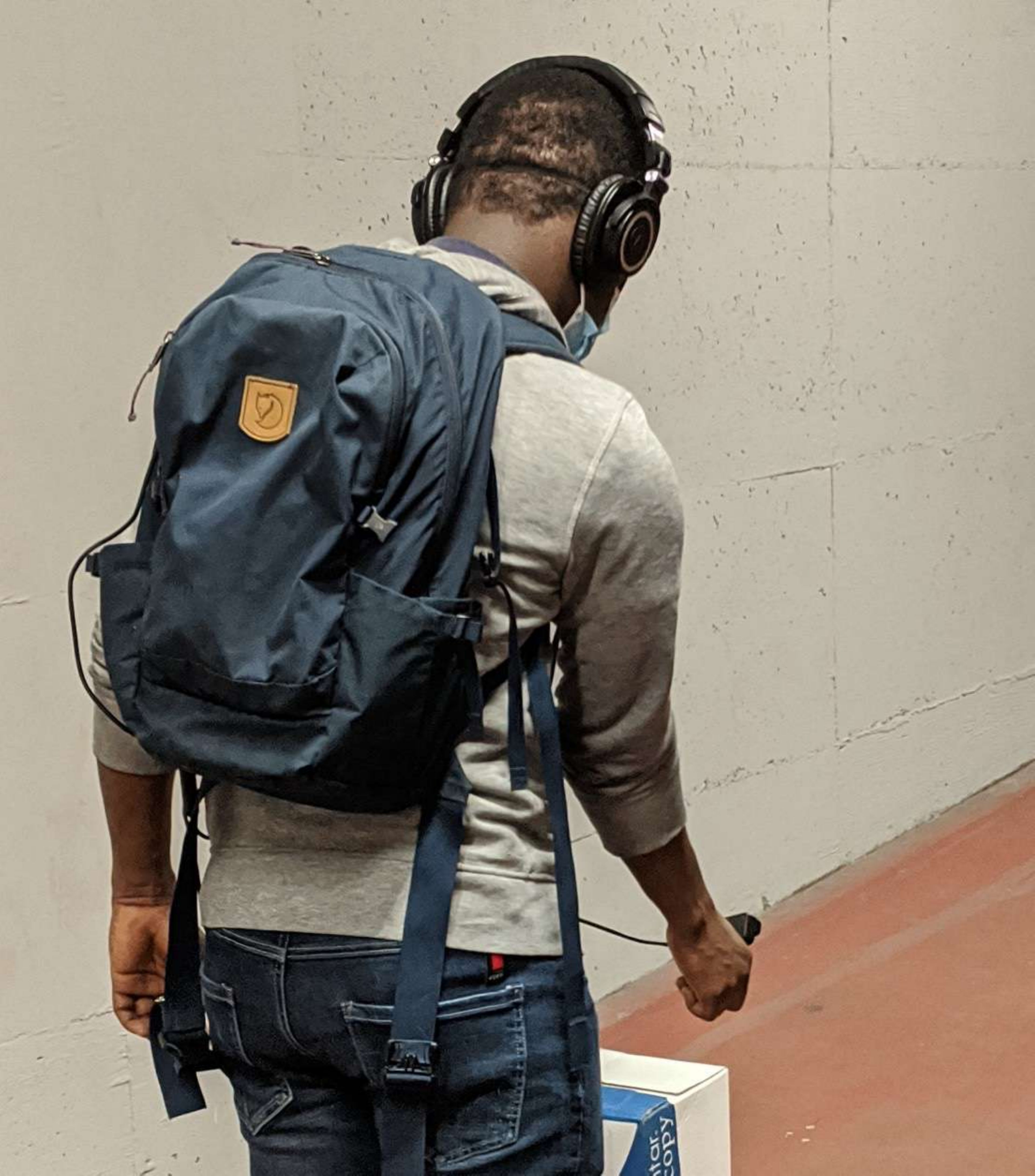} 
   \caption{Sensory substitution system worn by a participant. The camera and the headset are connected to the computer which is in the backpack worn by the participant. }
      \label{fig:expSetupPart}   
\end{figure}

The overall system consists of a handheld camera (\emph{Playstation Eye Camera}), a computer carried in a backpack and a headset plugged in the computer and worn by participants (Fig.~\ref{fig:expSetupPart}). 

\begin{figure}[htb] 
   \centering
   \includegraphics[width=0.7\textwidth]{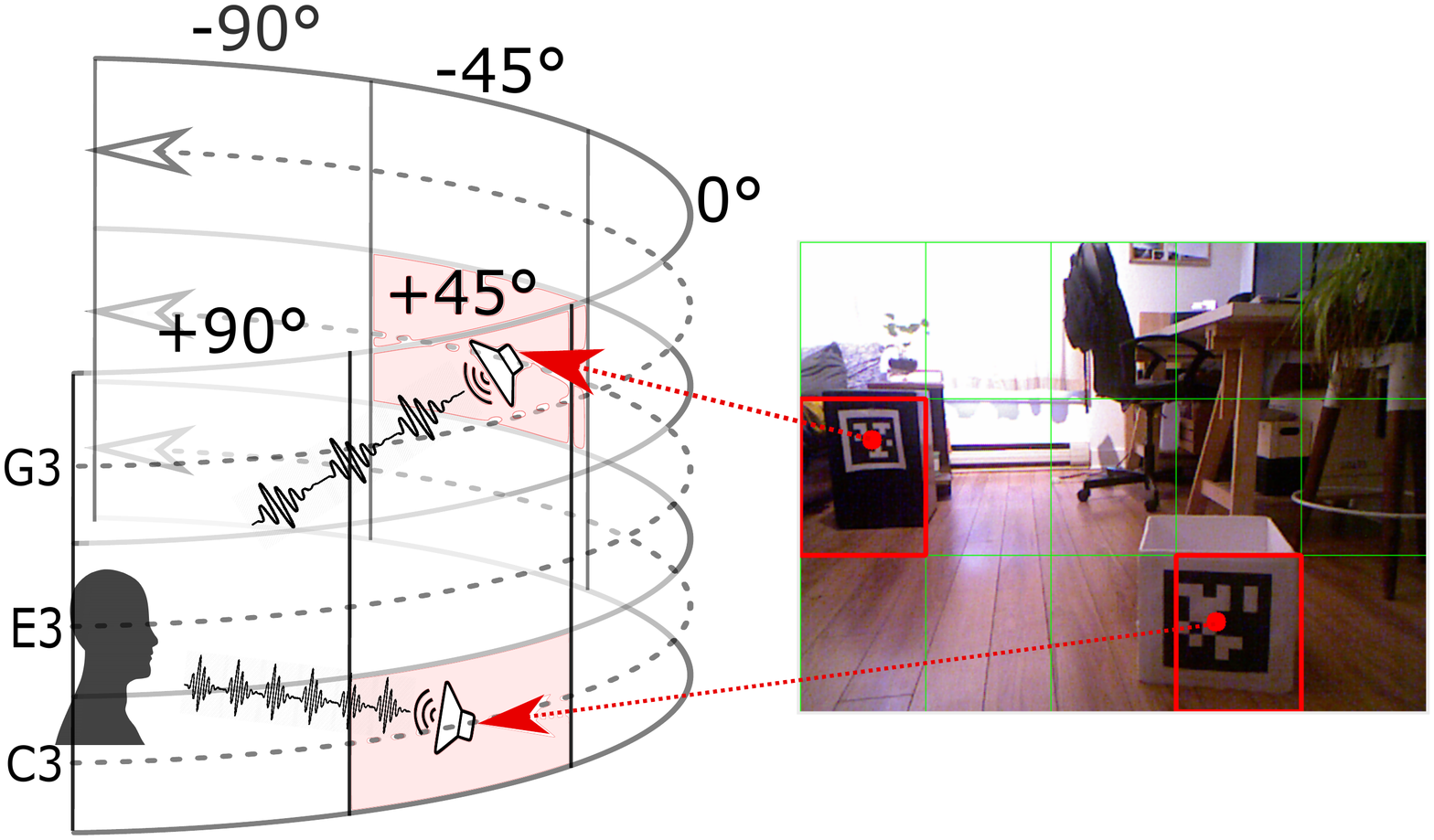}
   \caption{Principle of the modified version of See Differently~\cite{rouatICAD2014}. The 3D mode is illustrated here. When the object is far, the sound signal is repeated slowly (object on the left) while it is repeated quicker when the object is closer (object on the right). Three different musical notes are used to encode the rows of cells: C3 (bottom cells), E3 (middle cells) and G3 (top cells). These notes are spatialized with HRTFs filters to  encode the columns of cells.  
   }
      \label{fig:expSetupPhoto}   
\end{figure}

Figure~\ref{fig:expSetupPhoto} illustrates the principle of the substitution device. 
We adapted the 2D device~\textit{See Differenlty}~\cite{rouatICAD2014} 
to the needs of our study. The new device comprises 2D and 3D modes. Both encode in real-time the position of objects in images captured by a camera. In addition, with the 3D mode, the distance between the camera and the object is encoded.

We used the~\textit{AprilTag} algorithm~\cite{olson2011tags} to detect and estimate 3D relative positions of visual markers that we pasted on objects (Fig.~\ref{fig:expSetupPhoto}). 
With~\textit{AprilTag}, there is no need to use objects recognition algorithms and thus potential errors are avoided. 
The~\textit{AprilTag} algorithm is indeed robust since it uses known visual markers. The~\textit{AprilTag} markers were $4.3$ cm squares for the localization task and $17.3$ cm squares for the navigation task. 
Visual markers are detected at distances between 4 and 200 cm for the localization and between 14 and 900 cm for the navigation task. 

 The area of 2D images of the camera is divided into 15 rectangular cells. Each cell is associated with prerecorded spatialized piano sounds. 
 The horizontal position of the cells is encoded into the azimuth of the sounds (Fig.~\ref{fig:expSetupPhoto}). The spatialization is made via convolution with Head Related Transfer Function (HRTF) filters. We use the HRTF measured on the~\textit{KEMAR} dummy head from the CIPIC database~\cite{CIPIC2001}.
 To compensate for the poor elevation encoding with non-individual tunings of HRTF~\cite{stitt2019auditory}, we use pitches of piano notes to encode the vertical position of the cells which provides an estimate of the elevation of the objects. 
If a marker of an object is detected in a cell, its respective sound 
 is triggered and looped until the visual marker is no longer detected. There is no limit to the number of sound cells that can be played simultaneously. For the 2D mode, each spatialized piano note lasts 2 seconds. For the 3D mode, the duration of the note (in seconds) is equal to the distance (in meters) between the camera and the visual marker. For example, if the marker is $30$~cm from the camera the duration of the note will be $0.3$ seconds.

The visual marker recognition algorithm is implemented with~\textit{Python} and the sound synthesis is performed with the Supercollider environment~\cite{mccartney2002rethinking}.

\subsection{Protocol}
\label{sec:protocol}
The protocol was designed to compare the 2D and 3D modes of the system.
Experiments comprised two sessions, one for the 2D mode and one for the 3D mode.  
One session included a localization along with a navigation (three courses) task~\footnote{Both tasks were approved by the ethical committee from Letters and Human Sciences from Universit\'{e} de Sherbrooke under reference number 2014-85/Rouat.} with the same mode, either the 2D or the 3D (Table~\ref{tab:protocol}). 
 Participants were randomly split into two groups. 
Half of them completed their first experimental session with the 2D mode and their second experimental session with the 3D mode.
The other half completed the experimental sessions using the modes in the reverse order. This allowed us to study the impact of the order at which the 2D and 3D mode were used.
 
\begin{table}[htb]
\centering
\caption{Protocol. 
One localization task (Loc.) comprises the finding of 3 objects on a table. One navigation task comprises 3 courses. Participants in $Group_{2D3D}$ first completed the localization and the navigation with the 2D mode and then the 3D mode. Participants in $Group_{3D2D}$ completed the tasks using modes in the reverse order. \vspace{2mm}}
\begin{tabular}{|c|cc|cc|}
\hline
\multirow{2}{*}{Group} & \multicolumn{2}{c|}{Session 1}     & \multicolumn{2}{c|}{Session 2}     \\ \cline{2-5} 
                       & \multicolumn{1}{c|}{Loc.} & Navig. & \multicolumn{1}{c|}{Loc.} & Navig. \\ \hline
$Group_{2D3D}$            & \multicolumn{1}{c|}{2D}   & 2D     & \multicolumn{1}{c|}{3D}   & 3D     \\ \hline
$Group_{3D2D}$           & \multicolumn{1}{c|}{3D}   & 3D     & \multicolumn{1}{c|}{2D}   & 2D     \\ \hline
\end{tabular}
\label{tab:protocol}
\end{table}
%

%

\begin{figure}
\centering
\begin{subfigure}{.5\textwidth}
  \centering
  \includegraphics[width=.7\linewidth]{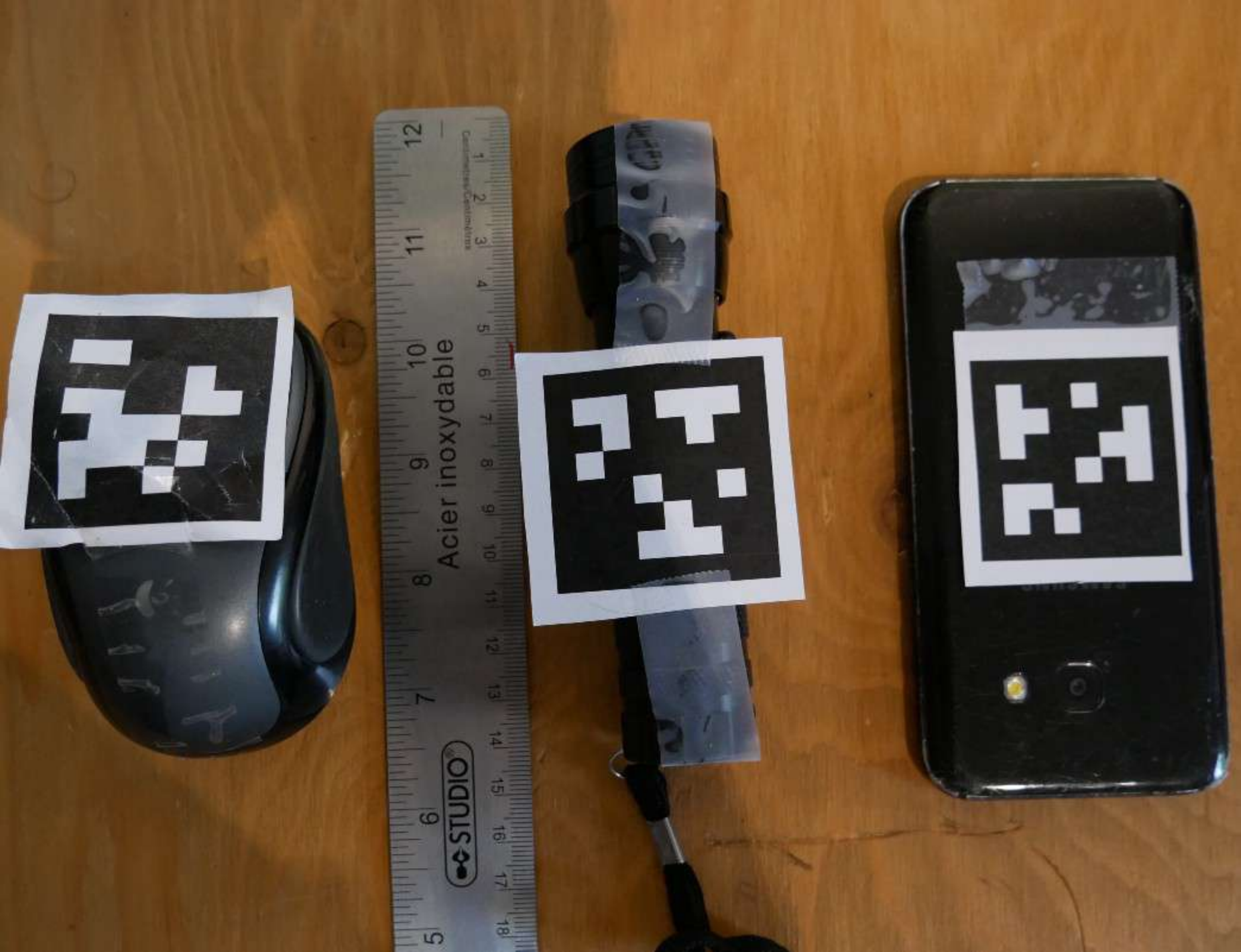}
  \caption{}
  \label{fig:sub1}
\end{subfigure}%
\begin{subfigure}{.5\textwidth}
  \centering
  \includegraphics[width=.7\linewidth]{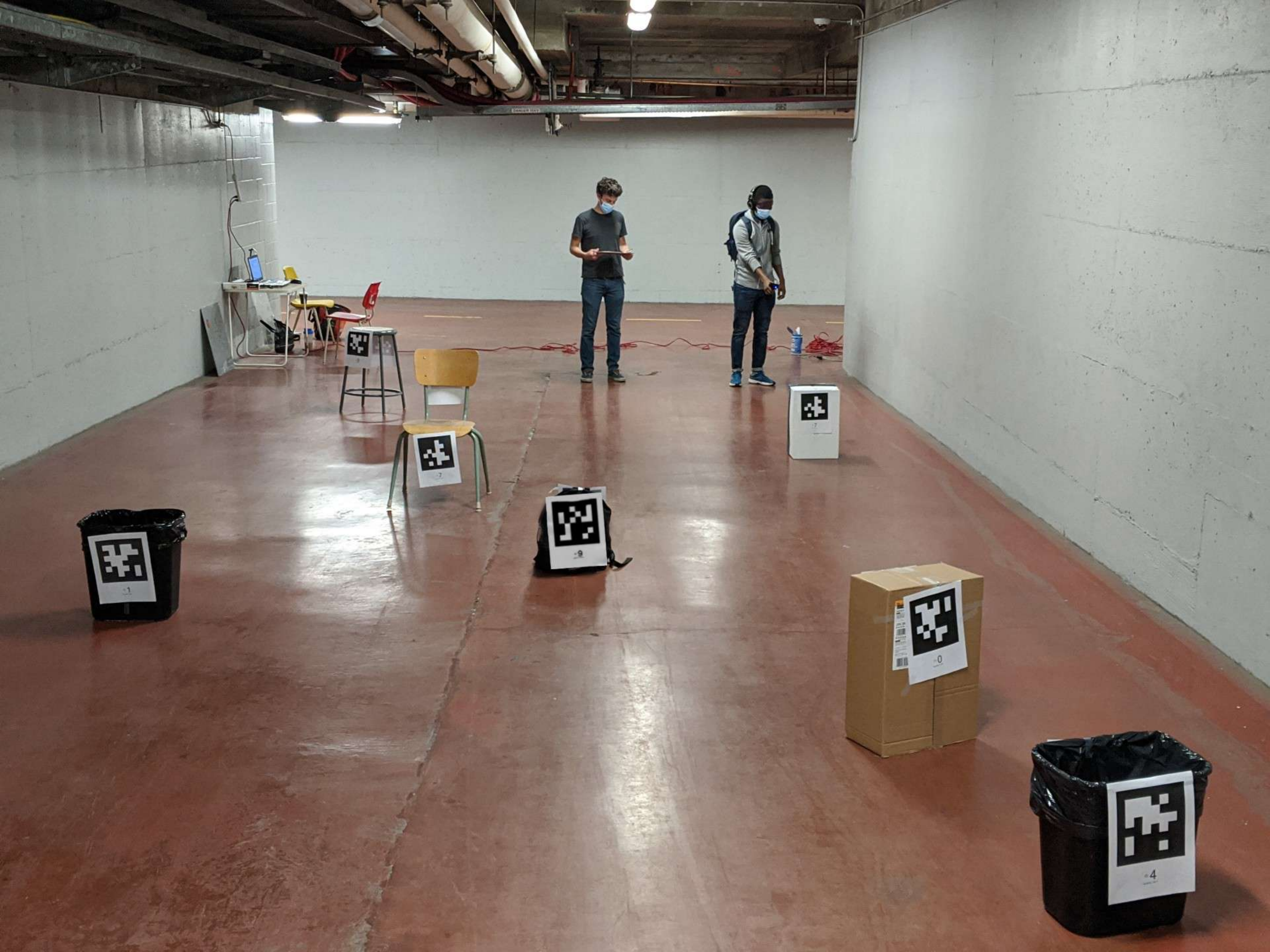}
  \caption{}
  \label{fig:sub2}
\end{subfigure}
\caption{Localization (a) and navigation (b). (a): A computer mouse, a flashlight, and a cellular phone have to be found on a table. (b) The participant scans the environment by moving the handheld device, while the assistant takes notes of the strategy being used by the participant.}
\label{fig:experiences}
\end{figure}

 The two experimental sessions, including the completion of the 2 tasks with the 2D and 3D mode, lasted on average between 30 minutes and 1 hour, depending on the time taken by  participants to complete the tasks. 
 Participants could either use their own headphones or the~\textit{Audio Technica} (model ATH-M50X) pair we provided. 
 They did not have any distinctive health-related problems. 
For the first task, participants had to localize 3 objects on a table (Fig.~\ref{fig:experiences}a). Once they became familiar with the device and located the 3 objects on the table, they could proceed to the navigation task. 



Participants had to fill up a questionnaire after completing the 2 experimental sessions with the 2D and 3D modes. Each participant received the same amount of financial compensation to cover their expenses and participation. None of them were relatives or friends of the authors.
 
\subsubsection{Localization task}
\label{sec:ProtocolLoc}
Participants were standing in front of a table, blindfolded, and were asked first to gradually increase the sound volume (set at the minimum before the experiment) to a comfortable level. They had to complete a short familiarization phase (5 to 10 minutes). During familiarization, they were allowed to explore and touch the table and the objects while 
listening to the sounds played by the device through headphones. Then, they were asked to locate three objects on the table: 
a wireless computer mouse, a mobile phone, and a flashlight (Fig.~\ref{fig:experiences} a). 
 They did not have to identify the objects. 
 We measured their location accuracy and speed in pointing at the 3 objects. 
  We proceeded as follows:
\begin{enumerate}[label=(\roman*)]
\item Participants are blindfolded and the 3 objects are placed randomly on the table within a range of 5 to 100 centimeters from participants;
\item Participants are asked to point in the direction of the first object they find;
\item Then, they need to touch with their index finger the object, 
or the table (in case of missing the object) so that we can measure the distance between the object and the index finger;
\item When the object is missed, the distance between the center of the object and the index finger of participants is labeled as the``localization error distance'';
\item Then, participants had to find the 2 other objects by following the same procedure from
 (iii) to (iv).
\end{enumerate}
Once an object is located, it is not removed from the table. The time taken to find each object and the distance to the object are recorded. 

\subsubsection{Navigation task}
\label{sec:ProtocolNavig}

Participants were asked to walk from one end of 
 the corridor (15 meters long and 6 meters wide) to the other without colliding with the objects from the scene (Fig.~\ref{fig:experiences} b). 
Two chairs, two garbage bins, two small bags, and two cardboard boxes were randomly placed in the corridor. Visual markers were pasted to the back and the front of each object. 
 Each participant had to make 
6 passes at their own pace: 3 passes with each mode (either 2D or 3D). 
 Participants were allowed to explore slowly or to go straight and fast.
 We proceeded as follows to measure participants' performance: 

\begin{enumerate}[label=(\roman*)]
\item Before each course, we blindfold participants
and randomly moved the obstacles. 
\item We start the stopwatch once participants are ready.
\item One assistant remains close to participants to take notes
about their strategy 
and to make sure that they do not collide with objects or walls. When participants detect an object on their way they have to stop walking and scanning and have to describe to the assistant where they think the object is located. The object is then reported by the assistant as ``seen'' if the description of the object location is accurate. 
Objects are classified as missed if the assistant has to momentarily interrupt the experiment to avoid a collision.
\item After the completion of a run, the time is noted.
\item Then, participants can rest or begin another run - restarting the process from step (i).
\item Once the 3 courses 
are completed, participants have to complete the second experimental session (starting from the localization) with the mode they have not yet used (either the 2D mode if they began with the 3D mode and vice versa). If this is already their second session, 
participants have to answer a questionnaire to give their feedback.
\end{enumerate}

\subsection{Graphical representation and analysis of results}
 Boxplots~\cite{rice2006mathematicalBoxplot} are used to show the distribution of localization errors (Fig.~\ref{fig:errorVsModeAndTrial}), 
 navigation times (Fig.~\ref{fig:courseTimeVsTrial}), and questionnaire responses (Fig.~\ref{fig:QuestionRating}). The horizontal lines of a boxplot are the first quartile (lower horizontal line), 
 the median (middle horizontal line), and third quartile (top line). The triangle is the average. 
 The ends of vertical lines are Minima and maxima (excluding outliers). 
 Data points that are more than 1.5 Interquartile range (IQR) away from the bottom or top quartile are outliers and are marked with black dots. 

We use different versions of the ANalysis Of Variance (ANOVA)~\cite{rice2006mathematicalANOVA} to analyze the effect of the variables in table~\ref{tab:protocol} (mode, group and session number) on participant's performance. An ANOVA determines whether the means of two or more  distributions are different, by comparing inter- and intra-group variances. For each ANOVA, we give the F statistic~\footnote{The F statistic represents the ratio between the inter- and the intra- group variance. F is computed from Fisher distribution and the degree of freedom of the inter- and intra- group.} and the p-value~\footnote{The p-value is computed from the F statistic and represent the probability of obtaining the observed means by chance.}. The significance level for the ANOVAs is set with the p-value at $p<0,05$. 

A One-factor repeated~\footnote{The repeated measure ANOVA is used when data are collected from the same participants under different conditions or at different times.} measure ANOVA is used to analyze the effect of the mode on localization (section~\ref{sec:loc}). A Two-factors~\footnote{The several factors ANOVA is used for analyzing the effect of several independent variables on one outcome variable}  repeated measure ANOVA  is used to evaluate the impact of the mode (2D or 3D) and the course number on navigation (section~\ref{sec:nav}). A two-factors ANOVA is also used to evaluate the effect of the mode and the group ($Group_{2D3D}$ or $Group_{3D2D}$) on the questionnaire (section~\ref{sec:quest}). 

\section{Results}

 \subsection{Localization}
 \label{sec:loc}
 Participants completed the localization task twice: once for each mode (2D and 3D). We study the impact of the the mode and the group ($Group_{2D3D}$ or $Group_{3D2D}$) . 

 The global average localization error (distance between participants' index finger and the center of objects) is $4.9\pm3.5$~cm. 

\begin{figure}[h] 
   \centering
   \includegraphics[width=0.5\textwidth]{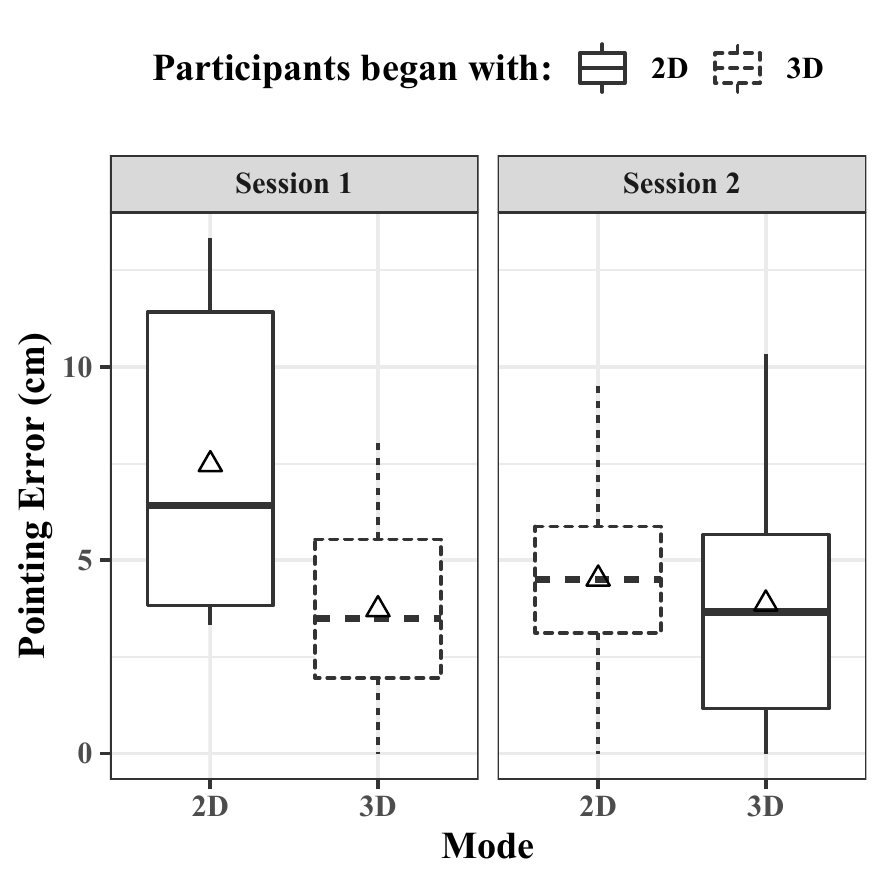}
   \caption{Localization task. Distributions of the participants' localization errors as a function of the session number (1 or 2) 
   and the system mode (2D or 3D). 
   Groups of participants who began with the 2D ($Group_{2D3D}$) and 3D ($Group_{3D2D}$) modes are shown with the continuous and dashed lines, respectively.
   }
      \label{fig:errorVsModeAndTrial}   
\end{figure}

Figure~\ref{fig:errorVsModeAndTrial} shows the distribution of the localization errors for each session number and each mode. To analyze the effect of the mode on the localization error, one-factor repeated measure ANOVAs were performed separately on each group. 
In the group of participants who began with the 2D mode ($Group_{2D3D}$), the analysis reveals that the mode (either 2D or 3D) has significant effect on localization errors ($F_{1,7} = 6.188,~p = \num{0.042}$). $Group_{2D3D}$ is significantly better with the 3D mode during their second session than with the 2D mode during their first session. In the group of participants who began with the 3D mode ($Group_{3D2D}$), the analysis shows no significant effect of the mode ($F_{1,7} = 0.279,~p = \num{0.614}$). $Group_{3D2D}$ obtains similar results at session 1 with 3D mode and session 2 with the 2D mode. 
Also, one-factor ANOVAs were computed per session (1 or 2) to study the effect of the group on localization errors. They show that the group has a significant impact during session 1 ($F_{1,14} = 4.796,~p = \num{0.046}$) but not during session 2 ($F_{1,14} = 0.167,~p = \num{0.167}$). 
Participants of $Group_{3D2D}$ who began with the 3D mode perform well in the first session. They are then able to create their strategy to accurately represent the 3D environment with the 2D mode during session 2. Participants of $Group_{2D3D}$ who began with the 2D mode are initially less accurate and perform significantly better during their second session with the 3D mode.  

Nevertheless, by taking into account the two experimental sessions, 11 out of 16 participants (69~\%) are more accurate with the 3D mode of the system.
 
Similar ANOVAs to those of the localization error were conducted to investigate the impact of mode and group on localization times. The analysis shows that neither the group nor the mode has a significant impact on the localization times. 
  
Finally, based on the speed-accuracy tradeoff consideration~\cite{heitz2014speed}, the hypothesis that participants could have adopted two different approaches to complete the task depending on their willingness was tested: either participant is quick but not accurate or accurate but slower. 
However, a Pearson's correlation coefficient shows that there is no relation between the time taken to complete the localization task and the localization error ($r(30) = 0.104,~p= 0.570$). 

\subsection{Navigation}
\label{sec:nav}
Participants completed 6 courses in total: three courses in a row with the 2D (or 3D) mode during session 1 and then three courses in a row with the 3D (or 2D) mode during session 2 (Table~\ref{tab:protocol}). We studied the impact of the mode, the group (i.e. order at which 2D and 3D were used) and the course number on performance.  

\begin{figure}[h] 
   \centering
   \includegraphics[width=0.5\textwidth]{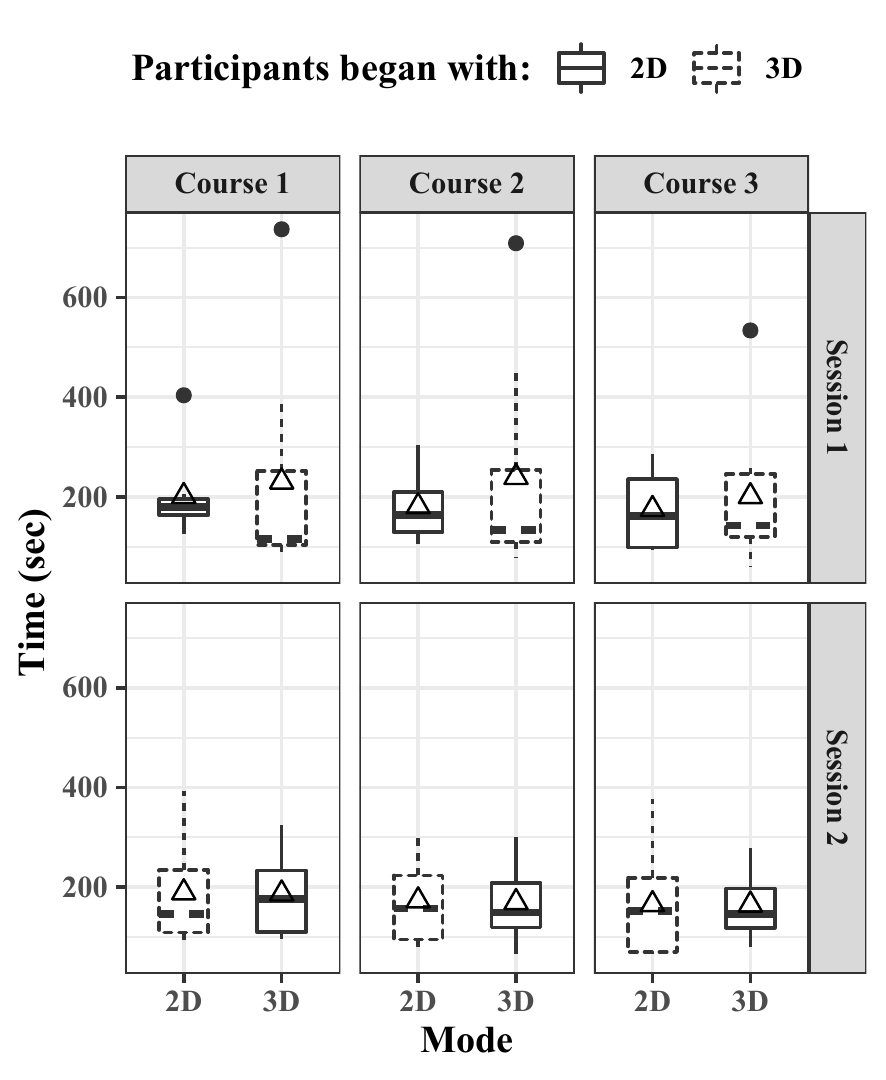}
   \caption{Navigation task. Distribution of the time taken by participants to complete the courses as a function of the session number, the mode and the course number. 
   For session one, some participants are considered to be outliers because of the difficulties they had to navigate. For session 2, we do not get outliers anymore. Averages (triangles) are lower for session 2.
   }
      \label{fig:courseTimeVsTrial}   
\end{figure}

Figure~\ref{fig:courseTimeVsTrial} shows the distribution of the time taken by the participants to complete one course as a function of the session number, the mode (2D or 3D) and the course number. Overall, Seven participants (44\%) achieved their best course time with the 2D mode and nine (56\%) with the 3D mode. 

To analyze the effect of the mode and the course number on the localization error, two-factors repeated measure ANOVAs were performed separately on $Group_{2D3D}$ that began with 2D and $Group_{3D2D}$ that began with 3D. For both groups, no significant effect of the mode ($Group_{2D3D}$: $F_{1,7}=1.335,~p=\num{0.286}$ , $Group_{3D2D}$: $F_{1,7}=1.939,~p=\num{0.206}$) or of the course numbers ($Group_{2D3D}$: $F_{2,14}=1.028,~p=\num{0.383}$, $Group_{3D2D}$: $F_{1.18,8.24}~\footnote{Non integer degrees of freedom for the F statistic are due to the Greenhouse Geisser correction applied to adjust for the lack of sphericity of data variances, which is a necessary assumption to conduct a repeated-measure ANOVA}=0.787,~p=\num{0.421}$) is found. Also, There is no interaction effect of the course number and the mode in $Group_{2D3D}$ ($F_{2,14}=0.015,~p=\num{0.985}$) and $Group_{3D2D}$ ($F_{1.09,7.62}=0.126,~p=\num{0.753}$).

Yet, no participants is considered an outliers for session 2 (with an average course time above 400 seconds) and average times decrease across courses and sessions (Fig.~\ref{fig:courseTimeVsTrial}). There is still a learning effect in the sense that performance becomes more consistent through time. 

A two-factors repeated measure ANOVAs was also performed on each group for the number of missed obstacles. As described in section~\ref{sec:ProtocolNavig}, we consider that an object is missed if the assistant had to momentarily interrupt the experiment to avoid a collision. The analysis shows no significant effect of the mode or the course number on the number of missed obstacles. All participants are able to complete the tasks with both modes by missing few objects (global average of $0.59\pm0.78$ missed obstacles).

As with localization, we hypothesized that participants could have been either fast and missed many objects or slow and missed few objects. 
As previously, a Pearson's correlation coefficient shows no relation between the time taken to complete the navigation and the number of missed objects ($r(30) = 0.021,~p= 0.835$). 

Regardless of the mode (2D or 3D), participants manage to efficiently use the system to detect obstacles and became more comfortable and faster. We report in the discussion (section~\ref{sec:oldExp}) results of complementary experiments conducted with 42 blindfolded sighted participants~\cite{commere2020evaluation}, who were using the initial 2D version of~\textit{See differently}. They are consistent with this work. 

\section{Questionnaire and qualitative observations}
\subsection{Description}

After the completion of the localization and navigation tasks, each participant was asked to complete a questionnaire (see Table~\ref{tab:questionsDeLouis}). 
Four types of questions were used: ``rating scale question'', 
``open-ended question'' and ``multiple choice question''. For ``rating scale questions'', participants were asked to provide a numerical response ranging from 1 to 5. For~\textit{[Understand., Loc. Ease, Navig. Ease]} questions, 1 and 5 meant ``very easy'' and ``very difficult'' respectively. For~\textit{[Navig. Afraid]}, 1 and 5 meant ``not at all afraid'' and ``very afraid'' respectively. 

The questionnaire was divided into three sections: one for the localization (questions~\textit{[Understanding,Loc. Ease, Loc. Strat.]}), one for the navigation (questions~\textit{[Navig. Ease, Navig. Afraid, Navig. Strat.]}) and one
to provide general feedback (~\textit{[Comfort]} question). 
Except for the~\textit{[Comfort]} question, participants were asked to answer the questions twice: once for each system mode.

\begin{table}[htbp]
   \centering
      \caption{List of questions that were answered after localization and navigation tasks. }

   \begin{tabular}{@{} llr @{}} 
      \toprule 
      Question & Type  \\
      \midrule
      \makecell[l]{Q1 - [Understand.] - How did you find the understanding of the  device's manipulation?}& Rating \\
      \makecell[l]{Q2 - [Loc. Ease] - How did you find the task of pointing the object with your index finger ?}& Rating \\
      Q3 - [Loc. Strat.] -  What was your strategy to find the objects on the table ? & Open \\
      Q4 - [Navig. Ease] -  How did you find the task of moving with the system ? & Rating \\
      Q5 - [Navig. Afraid] - Were you afraid to hit the obstacles ? & Rating \\
      Q6 - [Navig. Strat.] -  What was your strategy to move in the corridor ? & Open \\
      \makecell[l]{Q7 - [Comfort] - In general, were you more comfortable with or without the depth information?} & \makecell[l]{Multiple choice} \\
      \bottomrule 
   \end{tabular}
   \label{tab:questionsDeLouis}
\end{table}

\subsection{Results}
\label{sec:quest}
\begin{figure}[htb] 
   \centering
   \includegraphics[width=0.55\textwidth]{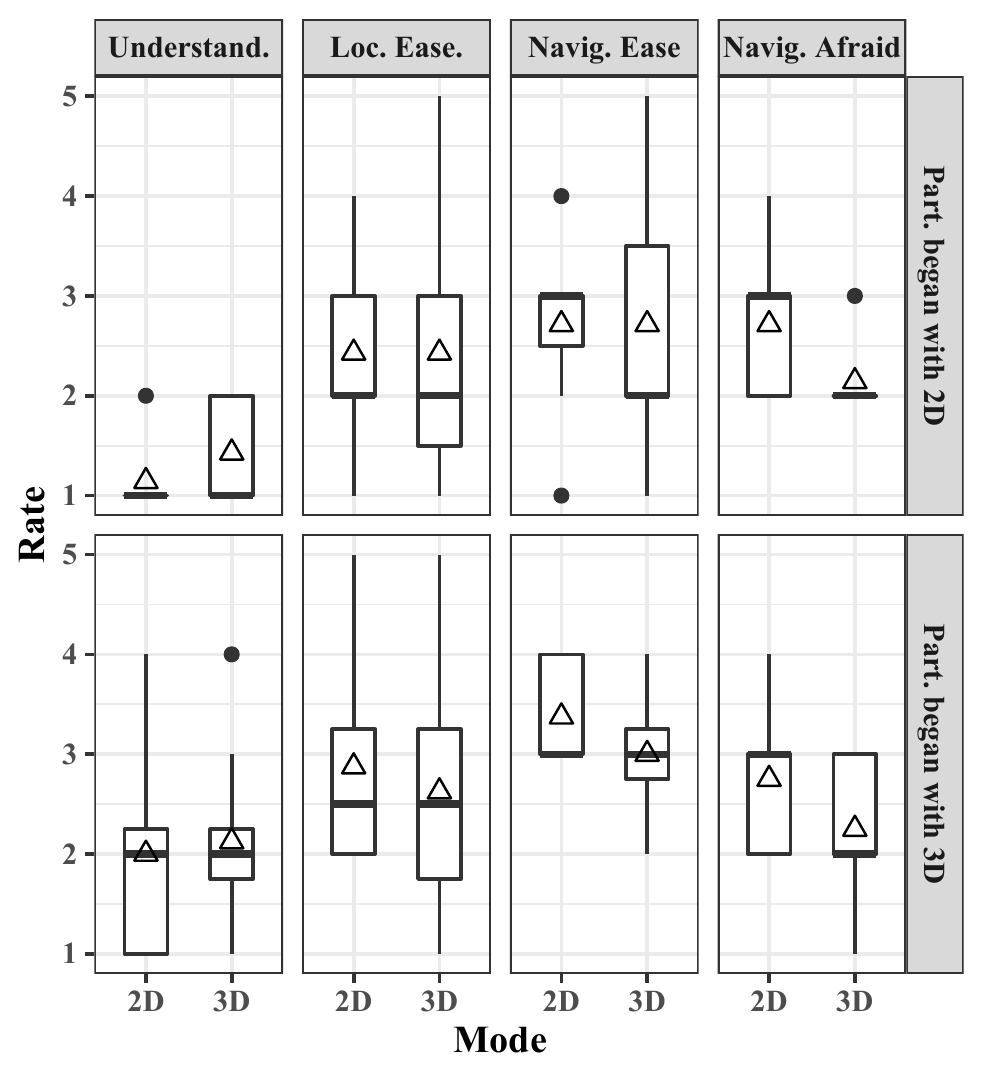}
   \caption{Questionnaire. Distribution of participants' ratings for the questions [Undestand., Loc. Ease, Navig. Ease, Navig. Afraid] as a function of the mode (2D or 3D) and the group (i.e. participants who began with 2D or 3D mode).}
      \label{fig:QuestionRating}   
\end{figure}

 Figure~\ref{fig:QuestionRating} shows the distribution of participants' answers to rating scale questions by mode and group.
  
 Two-factors ANOVAs were conducted for each rating question to study the impact of modes and the two groups on ratings given by participants. We report below significant ANOVAs. 
 For question [Understand.], the two-factors ANOVAs reveals a significant effect of the group  (i.e. order at which modes were used) on ratings ($F_{1,26}=6.714~,p=0.016$). Beginning with the 2D mode make it easier for participants to understand the system. Indeed, there is less information with the 2D mode (no audio encoded distance) which facilitate the interpretation of the sound. 
 For question [Navig. Afraid], the two-factors ANOVA reveals a significant effect of the mode on ratings ($F_{1,26}=4.915~,p=0.036$). Participants felt safer with the 3D mode, although they did not perform better during navigation.

Answers to questions [Loc. Strat.,Navig. Strat.] and observations made during the experiment revealed the strategies used for the localization and the navigation. 

\begin{figure}[htb] 
   \centering
   \includegraphics[width=0.5\textwidth]{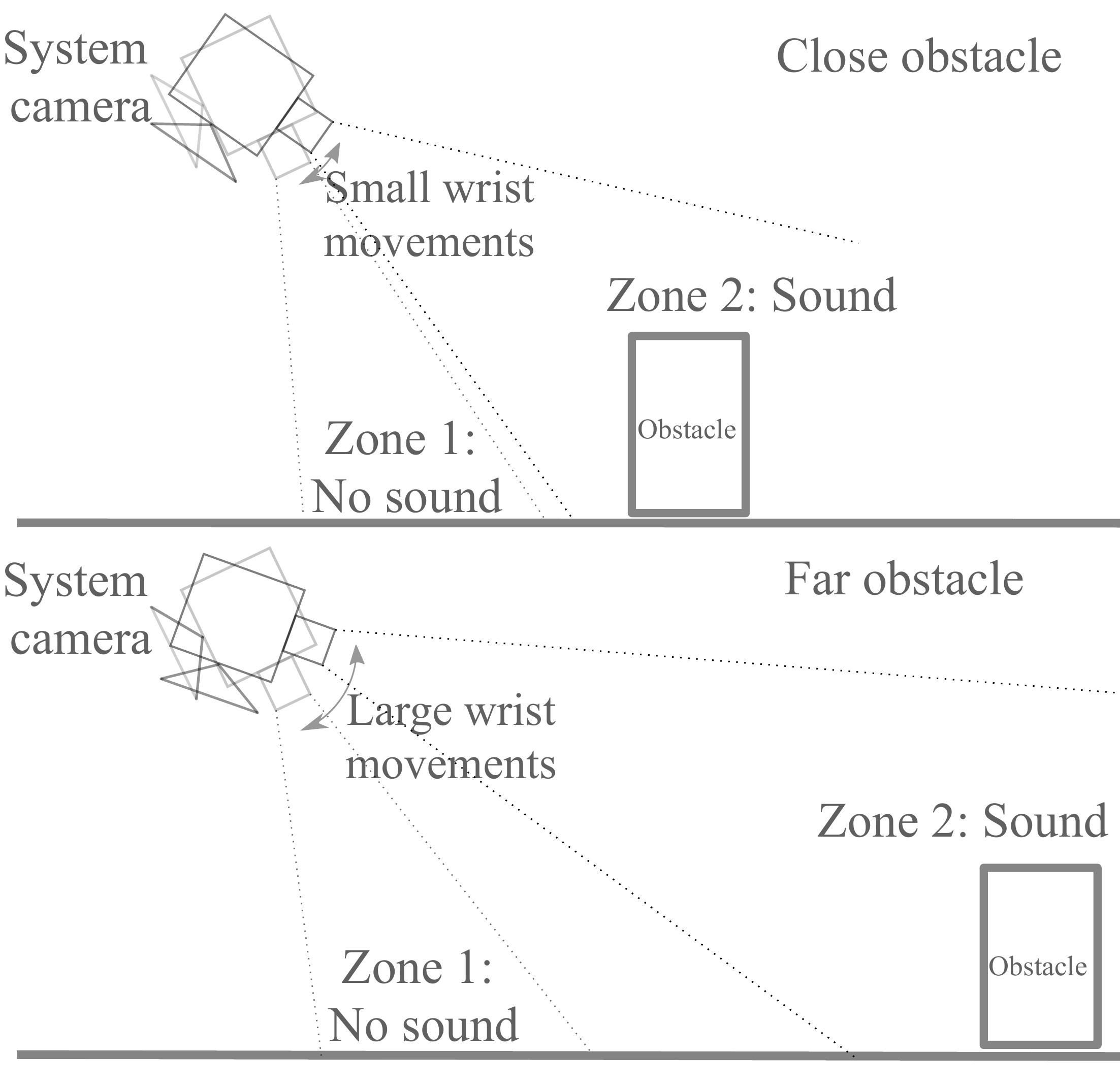}
   \caption{Navigation. Illustration of the main strategy used by participants to estimate the distance to an object. The distance is related to the up and down movement of the camera required to detect an obstacle. Top figure: the movement from zone 1 without obstacle (i.e. no sound) to zone 2 with obstacle (sound) is small which means that the object is close. Bottom figure: the movement from zone 1 to zone 2 is large which means that the object is far.}
\label{fig:depthEstimation2D}   
\end{figure}

For localization, a common and effective strategy is to scan first the entire surface of the table (i.e., hold the system high above the table) to get approximate localization. Then, the strategy is to get closer to localize more precisely an object. 
With the 3D mode, participants keep the sound centered~\footnote{i.e., keep the object on the cell that produces the E3 note with an azimuth angle of 0\degree, Fig.~\ref{fig:expSetupPhoto})} 
and use the sound rhythm that encodes the distance.  
With the 2D mode, participants estimate the distance by listening to the speed at which the sound changes when they move the device from right to left or up and down (see Fig.~\ref{fig:depthEstimation2D} for an``up and down'' movement example). When the movement required for the object to be detected in another cell (i.e. for the sound to change) is large, then the object is close, and the other way around. 

\begin{figure}[htb] 
   \centering
   \includegraphics[width=0.5\textwidth]{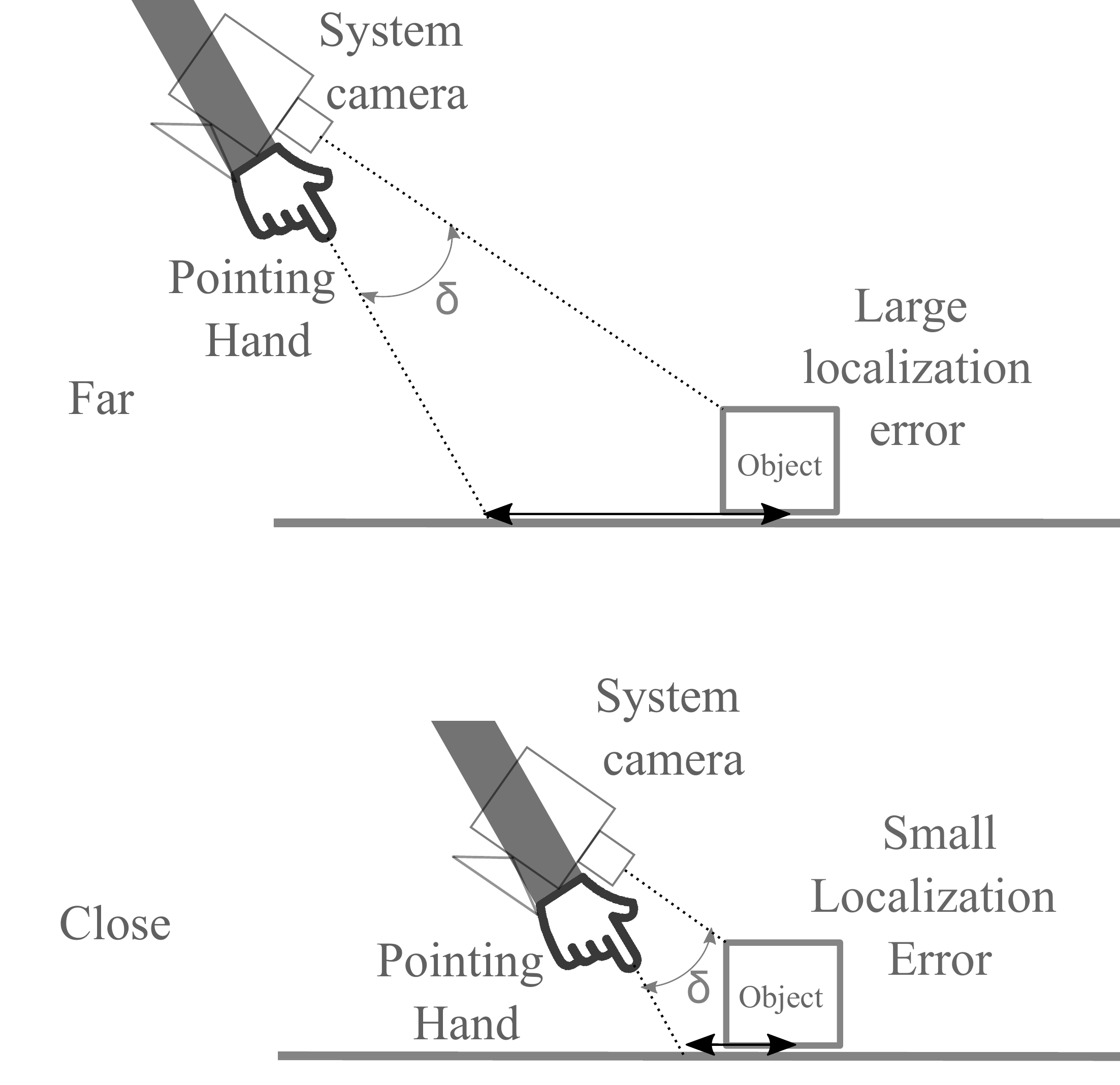}
   \caption{Localization task. Illustration of why localization errors are larger when participants try to point while the system is far from the object. The angle between the direction of the index finger and the camera is the same in both situations (close an far). When the camera is far from the object, the localization error is larger.
}
      \label{fig:locError}   
\end{figure}

We noticed that localization errors occurred primarily when participants attempted to point while holding the system away from the object. Indeed, the index finger of the pointing hand is often not oriented exactly in the same direction as the optical axis of the camera. When the camera is far from the object, the localization error is therefore more important (Fig.~\ref{fig:locError}).

For navigation, regardless of the mode, participants directed the camera towards the end of the corridor during the first course. 
The system was then detecting many obstacles on the course which made the sound cacophonous and difficult to interpret. Then, most  decided to scan right in front of them so that the device does not detect too many obstacles at the same time. 
Only 4 out of 16 participants reported using the audio encoded distance in the [Navig. Start] question compared to 12 out of 16 for localization in the [Loc. Start] question. They use a similar strategy to move and detect obstacles with both modes, as shown in Figure~\ref{fig:depthEstimation2D}.  



For the [Comfort] question, 13 out of 16 participants reported feeling more comfortable with the 3D mode. Among the 3 who felt more comfortable with the 2D mode, two (one in each group) found that the addition of the audio-encoded distance in the 3D mode overload their audition. 
A third reported feeling more comfortable with the 2D mode because he began the experiment with it. 

\section{Discussion} 

Early visual to auditory substitution devices encode 2D monocular images into sound while more recent systems increasingly use 3D cameras. However, no studies have investigated the impact of using audio encoded distance on the user experience. 
 The aim of the study was to quantify potential advantages of encoding distance in visual to auditory substitution devices. 
 This question is important to the design of new sensory substitution devices. 
 We therefore conducted experiments involving object localization and navigation tasks with a handheld visual to auditory substitution system. The system comprises 2 modes. Both modes encode the position of objects in 2D images and one of the two mode additionally encodes the distance between the system and the object.
The experiments were conducted with 16 blindfolded sighted participants. 

\subsection{Consistency of the results with prior experiments}
\label{sec:oldExp}
In~\cite{commere2020evaluation}, we conducted prior experiments with forty-two blindfolded sighted participants 
who were using the initial 2D version of the~\textit{See Differently} substitution device~\cite{rouatICAD2014}. This initial version used a neural-network based image processing algorithm to detect objects. 
As in this work, the system was encoding positions of objects in 2D images, by triggering sounds associated to the cells. There were 12 cells (3 rows, 4 columns) in that initial version and it did not provide distance information. 
Participants had to perform the same localization and navigation tasks.  
 We used the same protocols as described in section~\ref{sec:ProtocolLoc} (localization) and~\ref{sec:ProtocolNavig} (navigation) of this paper. We also used the same objects for localization (Fig.~\ref{fig:experiences} a) and the same obstacles for navigation (Fig.~\ref{fig:experiences} b) in the same corridor. 

For the navigation task, we observed a  learning effect across the 5 courses (average time decreased between the first and last trials of $107\pm78$ seconds) and participants missed very few objects ($0.78\pm0.87$ ). As for this work, the number of missed objects did not decrease significantly between courses. 
Overall, the results presented here with the 2D mode are consistent with those from experiments from~\cite{commere2020evaluation} that were conducted with 42 participants. We are therefore confident about the generalization of the results presented in this paper. 

\subsection{The handheld 2D device allows to perceive 3D environment}
\label{sec:2DPerception}
In line with studies of Ward~\textit{et al.}~\cite{ward2010}, Auvray~\textit{et  al.}~\cite{auvray2007learning} with~\textit{the vOICe}, and the study of Renier and Volder with the~\textit{PSVA}~\cite{Renier2010}, our participants were able to perceive the 3D environment with 
 the encoding of 2D monocular images into sounds. 
 They inferred distance with wrist movements as illustrated in Figure~\ref{fig:depthEstimation2D}. It suggests that participants are able to assess the amplitude of their wrist movements and relate it to the sound changes. 
 
 The ability to perceive the position of a joint without visual  feedback or other external cues is named ``joint position sense'' (or joint proprioception).
Wrist position sense has been widely studied
~\cite{gay2010new,marini2016wrist,pilbeam2018test,li2019wrist}. Although it depends on the protocols, measurement systems and the geometrical axis in which the joint is studied, the accuracy of wrist position sense has always found to be below 10\degree. Marini~\textit{et al.}~\cite{marini2016wrist} found that the wrist position sense allows the assessment of the difference between two positions ("vector coding") rather than an absolute position ("positional coding"). 
This supports the fact that our participants were able to accurately estimate the wrist movement required to direct the camera from zone 1 with no sound to zone 2 with sound (Fig.~\ref{fig:depthEstimation2D}). This allows them to evaluate the distance of objects without the need for explicit distance encoding. 


\subsection{The 3D handheld device allows faster learning}
For object localization, an effective strategy is to move the device as close as possible to the targeted object before pointing (Fig.~\ref{fig:locError}). The quantitative results show that the 3D mode is at first more effective than the 2D mode in providing small localization errors. 
The strategy is easier and faster to implement with the 3D mode thanks to the distance encoded by the sound rhythm. After practicing, the 2D mode allows participants to get close to the object as well and thus make small localization errors. It suggests that sonification of 2D monocular images allows to perceive the 3D environment after some training. On the other hand, the explicit sonification of distances allows the perception of the 3D environment without prior learning. 

As reported in the literature, ``2D systems'' often require more training time than ``3D systems'' to perform spatial tasks. Participants using~\textit{the vOICe} and the~\textit{PSVA} (which are``2D systems'') needed to train for 5 and 15 hours~\cite{auvray2007learning,Renier2010} before successfully completing the spatial tasks. When performing spatial tasks with 3D systems~\cite{Brock2013,stoll2015navigating,Skulimowski2018}, participants often only need to know how the system works.







\subsection{Do we need explicit audio encoded distances?}
Qualitative results show that most participants felt more comfortable with the mode that explicitly encodes distances. Indeed, the 3D mode initially allows for better localization accuracy. In addition, participants felt more secure with the 3D mode during the navigation. 
Although most participants did not exploit the audio-encoded distance of the 3D mode, they felt that they could better anticipate obstacles avoidance.

There are still 18\% of participants who were more comfortable without the audio encoded distance. They claimed that it overloaded their audition.  Also, for navigation, participants used a similar strategy with the 2D and 3D modes that did not require distance information (Fig.~\ref{fig:depthEstimation2D}). This explains why we did not observe differences between the two modes during this task. 
Therefore, audio-encoded distance did not improve participants' performance during the navigation. 

On one hand, the 3D mode gives equivalent or better quantitative and qualitative results. On the other hand, practice allowed users of the 2D mode to compensate for the lack of distance encoding during localization. During navigation, performance was similar with and without the explicit encoding of distance. In addition, although they are increasingly accessible, 3D sensors are still more expensive. Overall, we support the idea of Dakopoulos and Bourbakis that an SSD should be ``easy to use (operation and interface not loaded with unnecessary features) and without the need of an extensive training period''~\cite{Dakopoulos2010}. Thus, removing unnecessary audio-encoded information can make the interpretation of the sound easier for some people. 

 Since we have shown that the effectiveness of audio encoded distance depends on the task and people's preferences, we believe that systems should give users the ability to choose which information (especially distance information) will be encoded in sound based on their preferences and on usage contexts. 

\subsection{Limitations of the work}
 We chose to investigate handheld systems rather than a head-mounted. As suggested by Auvray~\textit{et al.}~\cite{auvray2007learning} handheld systems give the user more freedom to explore the visual scene. In addition, a handheld system can be implemented in a smartphone or similar device. Since nowadays most people have a smartphone, the system would be more discreet in public environments. 
However, it is still interesting to conduct experiments with head-mounted devices. 
  For example, Brown~\textit{et al.}~\cite{Brown2011} showed that~\textit{the vOICe} substitution system was more efficient at identifying objects when held in the hand while it was more efficient at navigating when mounted on the head. 
  
  In this work we use the same encoding of 2D characteristics of the visual scenes for a more objective comparison of the 2D and 3D modes. In this context, the 3D mode does not bring much improvement over the 2D mode when the participant is well trained. 
An interesting strategy that we leave as a future work would be to use the distance information to segregate the closest objects in the scene from the objects that are far. During navigation, this would have made the sound less confusing when participants were pointing the camera towards the end of the corridor. Participants could then have adopted different strategies with the 2D and 3D modes. 
  
  
Finally, experiments have been conducted with blindfolded but sighted individuals. 
Indeed, the results should be similar with late-blind participants since they adopt similar strategies to those used by sighted individuals when performing spatial tasks~\cite{ungar2000,Pasqualotto2013}. As a future work, it would be interesting to compare our results with further experiments done with late blind individuals and eventually congenitally blind.

\section{Conclusion}
Vision to audition substitution devices are very promising for blind or low vision people. 
 In this work, we evaluated potential advantages of sonified distances for spatial tasks. 
Overall, the 3D mode that encodes distances into sounds yield comparable performance to the 2D mode. 
Indeed, we show that the use of audio encoded distance allows users to understand 3D scenes with less learning compared to not using distance. Users are also more confident and feel more secure when the distance is explicitely encoded. However, results should be nuanced. Without audio encoded distance, users learn to develop efficient strategies to perceive the 3D environment. Also, depending on the type of  task, explicit distance encoding does not necessary yield better results. Finally, some participants prefer the 2D mode which does not encode distance. 
 Ideally, we would recommend to the designers of vision to audition substitution systems to let the users decide which information to sonify, depending on their preference and tasks.

\section*{Acknowledgment}

D. Lescal for a preliminary implementation and design of See Differently. FRQNT and NSERC-CRSNG for funding
this research.  A. Yarga and E. Calvet for the assistance during experiments. L. Celotti and S. Wood and the members of the NECOTIS research group for proofreading the paper. The participants, E. Plourde and
the members of the NECOTIS research group for testing
and providing feedbacks. Franco Lepore and Patrice Voss for
stimulating and fruitful discussions for the planning of the
 protocol. Fran\c cois C\^ ot\' e for the fruitful discussions on the blind community.

\bibliographystyle{unsrt}
\bibliography{template}  

\begin{thebibliography}{10}

\bibitem{bach1987brain}
Paul Bach-y{-}Rita.
\newblock Brain plasticity as a basis of sensory substitution.
\newblock {\em Journal of Neurologic Rehabilitation}, 1(2):67--71, 1987.

\bibitem{kokovayNeuron2008}
Erzsebet Kokovay, Qin Shen, and Sally Temple.
\newblock Perspective the incredible elastic brain : How neural stem cells
  expand our minds.
\newblock {\em Neuron}, 60(3):420--429, 2008.

\bibitem{bachYRita1969}
Paul Bach-y{-}Rita, C.C. Collins, F.~Saunders, B.~White, and L.~Scadden.
\newblock Vision substition by tactile image projection.
\newblock {\em Nature}, 221:963--964, 1969.

\bibitem{bachYRitaTrends2003}
P.~Bach-y{-}Rita and SW. Kercel.
\newblock Sensory substitution and the human-machine interface.
\newblock {\em Trends Cogn Sci}, 7:541--546, 2003.

\bibitem{ptitoBrain2005}
Maurice Ptito, Solvej~M. Moesgaard, Albert Gjedde, and Ron Kupers.
\newblock Cross-modal plasticity revealed by electrotactile stimulation of the
  tongue in the congenitally blind.
\newblock {\em Brain}, 128(3):606--614, 2005.

\bibitem{csapo2015survey}
{\'A}d{\'a}m Csap{\'o}, Gy{\"o}rgy Wers{\'e}nyi, Hunor Nagy, and Tony Stockman.
\newblock A survey of assistive technologies and applications for blind users
  on mobile platforms: a review and foundation for research.
\newblock {\em Journal on Multimodal User Interfaces}, 9(4):275--286, 2015.

\bibitem{palmer1999vision}
Stephen~E Palmer.
\newblock {\em Vision science: Photons to phenomenology}.
\newblock The MIT press, 1999.

\bibitem{Meijer1992}
P.~B.~L. Meijer.
\newblock An experimental system for auditory image representations.
\newblock {\em IEEE Transactions on Biomedical Engineering}, 39(2):112--121,
  Feb 1992.

\bibitem{Capelle1998}
C.~Capelle, C.~Trullemans, P.~Arno, and C.~Veraart.
\newblock A real-time experimental prototype for enhancement of vision
  rehabilitation using auditory substitution.
\newblock {\em IEEE Transactions on Biomedical Engineering}, 45(10):1279--1293,
  Oct 1998.

\bibitem{Auvray2006}
Malika Auvray, Sylvain Hanneton, Charles Lenay, and Kevin O'Regan.
\newblock There is something out there: Distal attribution in sensory
  substitution, twenty years later.
\newblock {\em Journal of integrative neuroscience}, 4:505--21, 2005.

\bibitem{abboud2014eyemusic}
Sami Abboud, Shlomi Hanassy, Shelly Levy-Tzedek, Shachar Maidenbaum, and Amir
  Amedi.
\newblock Eyemusic: Introducing a ÒvisualÓ colorful experience for the blind
  using auditory sensory substitution.
\newblock {\em Restorative neurology and neuroscience}, 32(2):247--257, 2014.

\bibitem{ambard2015mobile}
Maxime Ambard, Yannick Benezeth, and Philippe Pfister.
\newblock Mobile video-to-audio transducer and motion detection for sensory
  substitution.
\newblock {\em Frontiers in ICT}, 2:20, 2015.

\bibitem{ward2010}
J.~Ward and P.~B.~L. Meijer.
\newblock Visual experiences in the blind induced by an auditory sensory
  substitution device.
\newblock {\em Consciousness and cognition}, 19(1):492--500, 2010.

\bibitem{auvray2007learning}
Malika Auvray, Sylvain Hanneton, and Kevin O{'}Regan.
\newblock Learning to perceive with a visuo-auditory substitution system:
  Localisation and object recognition with 'the v{OIC}e'.
\newblock {\em Perception}, 36(3):416--430, 2007.

\bibitem{Renier2010}
Laurent Renier and A.~G. {De Volder}.
\newblock Vision substitution and depth perception: Early blind subjects
  experience visual perspective through their ears.
\newblock {\em Disability and Rehabilitation: Assistive Technology},
  5(3):175--183, 2010.

\bibitem{Brock2013}
Michael Brock and Per~Ola Kristensson.
\newblock Supporting blind navigation using depth sensing and sonification.
\newblock In {\em Proceedings of the 2013 ACM Conference on Pervasive and
  Ubiquitous Computing Adjunct Publication}, UbiComp '13 Adjunct, page
  255Ð258, New York, NY, USA, 2013. Association for Computing Machinery.

\bibitem{stoll2015navigating}
Chlo{\'e} Stoll, Richard Palluel-Germain, Vincent Fristot, Denis Pellerin,
  David Alleysson, and Christian Graff.
\newblock Navigating from a depth image converted into sound.
\newblock {\em Applied bionics and biomechanics}, 2015, 2015.

\bibitem{Skulimowski2018}
Piotr Skulimowski, Mateusz Owczarek, Andrzej Radecki, Michal Bujacz, Dariusz
  Rzeszotarski, and Pawel Strumillo.
\newblock Interactive sonification of u-depth images in a navigation aid for
  the visually impaired.
\newblock {\em Journal on Multimodal User Interfaces}, Nov 2018.

\bibitem{yang2016expanding}
Kailun Yang, Kaiwei Wang, Weijian Hu, and Jian Bai.
\newblock Expanding the detection of traversable area with realsense for the
  visually impaired.
\newblock {\em Sensors}, 16(11):1954, 2016.

\bibitem{yang2018long}
Kailun Yang, Kaiwei Wang, Shufei Lin, Jian Bai, Luis~M Bergasa, and Roberto
  Arroyo.
\newblock Long-range traversability awareness and low-lying obstacle
  negotiation with realsense for the visually impaired.
\newblock In {\em Proceedings of the 2018 International Conference on
  Information Science and System}, pages 137--141, 2018.

\bibitem{aladren2016navigation}
A.~{Aladr?n}, G.~{LÑpez-Nicolàs}, L.~{Puig}, and J.~J. {Guerrero}.
\newblock Navigation assistance for the visually impaired using rgb-d sensor
  with range expansion.
\newblock {\em IEEE Systems Journal}, 10(3):922--932, Sep 2016.

\bibitem{Kayukawa2019}
Seita Kayukawa, Keita Higuchi, J.~Guerreiro, Shigeo Morishima, Yoichi Sato,
  Kris Kitani, and Chieko Asakawa.
\newblock B{B}eep: A sonic collision avoidance system for blind travellers and
  nearby pedestrians.
\newblock In {\em Proceedings of the 2019 CHI Conference}, page 1Ð12, New
  York, NY, USA, 2019. Association for Computing Machinery.

\bibitem{Bai2019}
Jinqiang Bai, Zhaoxiang Liu, Yimin Lin, Ye~Li, Shiguo Lian, and Dijun Liu.
\newblock Wearable travel aid for environment perception and navigation of
  visually impaired people.
\newblock {\em Electronics}, 8(6), 2019.

\bibitem{Li2020}
Z.~{Li}, F.~{Song}, B.~C. {Clark}, D.~R. {Grooms}, and C.~{Liu}.
\newblock A wearable device for indoor imminent danger detection and avoidance
  with region-based ground segmentation.
\newblock {\em IEEE Access}, 8:184808--184821, 2020.

\bibitem{kristjansson2016designing}
{\'A}rni Kristj{\'a}nsson, Alin Moldoveanu, {\'O}mar~I J{\'o}hannesson, Oana
  Balan, Simone Spagnol, Vigd{\'\i}s~Vala Valgeirsd{\'o}ttir, and R{\'u}nar
  Unnthorsson.
\newblock Designing sensory-substitution devices: Principles, pitfalls and
  potential.
\newblock {\em Restorative neurology and neuroscience}, 34(5):769--787, 2016.

\bibitem{rouatICAD2014}
Jean Rouat, Damien Lescal, and Sean Wood.
\newblock Handheld device for substitution from vision to audition.
\newblock In {\em the 20th Int. Conf. on Auditory Display}, New--York city,
  June 2014.

\bibitem{olson2011tags}
Edwin Olson.
\newblock Apriltag: A robust and flexible visual fiducial system.
\newblock In {\em Proceedings of the {IEEE} International Conference on
  Robotics and Automation ({ICRA})}, pages 3400--3407. IEEE, May 2011.

\bibitem{CIPIC2001}
V.R. Algazi, R.O. Duda, D.M. Thompson, and C.~Avendano.
\newblock The {CIPIC} {HRTF} database.
\newblock In {\em Proceedings of the 2001 IEEE Workshop on the Applications of
  Signal Processing to Audio and Acoustics}, pages 99--102, 2001.

\bibitem{stitt2019auditory}
Peter Stitt, Lorenzo Picinali, and Brian~FG Katz.
\newblock Auditory accommodation to poorly matched non-individual spectral
  localization cues through active learning.
\newblock {\em Scientific reports}, 9(1):1063, 2019.

\bibitem{mccartney2002rethinking}
James McCartney.
\newblock Rethinking the computer music language: Supercollider.
\newblock {\em Computer Music Journal}, 26(4):61--68, 2002.

\bibitem{rice2006mathematicalBoxplot}
John~A Rice.
\newblock {\em Mathematical statistics and data analysis}, chapter Chapter 10:
  Summarizing Data, pages 402--404.
\newblock Thomson Higher Education, 2006.

\bibitem{rice2006mathematicalANOVA}
John~A Rice.
\newblock {\em Mathematical statistics and data analysis}, chapter Chapter 12:
  The Analysis of Variance, pages 477--513.
\newblock Thomson Higher Education, 2006.

\bibitem{heitz2014speed}
Richard~P Heitz.
\newblock The speed-accuracy tradeoff: history, physiology, methodology, and
  behavior.
\newblock {\em Frontiers in neuroscience}, 8:150, 2014.

\bibitem{commere2020evaluation}
Louis Comm\`ere, Sean~UN Wood, and Jean Rouat.
\newblock Evaluation of a vision-to-audition substitution system that provides
  {2D} {WHERE} information and fast user learning.
\newblock {\em arXiv preprint arXiv:2010.09041}, 2020.

\bibitem{gay2010new}
Andre Gay, Kimberly Harbst, Kenton~R Kaufman, Diana~K Hansen, Edward~R
  Laskowski, and Richard~A Berger.
\newblock New method of measuring wrist joint position sense avoiding cutaneous
  and visual inputs.
\newblock {\em Journal of neuroengineering and rehabilitation}, 7(1):1--7,
  2010.

\bibitem{marini2016wrist}
Francesca Marini, Valentina Squeri, Pietro Morasso, and Lorenzo Masia.
\newblock Wrist proprioception: amplitude or position coding?
\newblock {\em Frontiers in neurorobotics}, 10:13, 2016.

\bibitem{pilbeam2018test}
Chlo{\"e} Pilbeam and Victoria Hood-Moore.
\newblock Test--retest reliability of wrist joint position sense in healthy
  adults in a clinical setting.
\newblock {\em Hand Therapy}, 23(3):100--109, 2018.

\bibitem{li2019wrist}
Lin Li, ShuWang Li, and YanXia Li.
\newblock Wrist joint proprioceptive acuity assessment using inertial and
  magnetic measurement systems.
\newblock {\em International Journal of Distributed Sensor Networks}, 15(4),
  2019.

\bibitem{Dakopoulos2010}
D.~Dakopoulos and N.~G. Bourbakis.
\newblock Wearable obstacle avoidance electronic travel aids for blind: A
  survey.
\newblock {\em IEEE Transactions on Systems, Man, and Cybernetics, Part C
  (Applications and Reviews)}, 40(1):25--35, Jan 2010.

\bibitem{Brown2011}
David Brown, Tom Macpherson, and Jamie Ward.
\newblock Seeing with sound? exploring different characteristics of a
  visual-to-auditory sensory substitution device.
\newblock {\em Perception}, 40(9):1120--1135, 2011.

\bibitem{ungar2000}
Simon Ungar.
\newblock Cognitive mapping without visual experience.
\newblock {\em Cognitive mapping: past, present, and future}, 4:221, 2000.

\bibitem{Pasqualotto2013}
Achille Pasqualotto, Mary~Jane Spiller, Ashok~S Jansari, and Michael~J Proulx.
\newblock Visual experience facilitates allocentric spatial representation.
\newblock {\em Behavioural Brain Research}, 236:175 -- 179, 2013.

\end{thebibliography}






\end{document}